\documentclass[a4paper, 11pt]{article}
\usepackage{cite}
\usepackage{pos}
\usepackage{physics}
\usepackage[alsoload=hep]{siunitx}
\usepackage{url}
\usepackage[frozencache,cachedir=minted-files]{minted}
\usepackage{wrapfig}
\usepackage{subcaption}
\usepackage{graphicx}

\usepackage{color}
\usepackage{tikz}
\usetikzlibrary{decorations.pathmorphing,decorations.markings}
\usetikzlibrary{decorations.pathreplacing}
\usetikzlibrary{positioning, calc, shapes}
\usetikzlibrary{shapes.geometric, arrows, calc}
\tikzstyle{tool} = [rectangle, rounded corners, minimum width=1cm, text
					centered, draw=black, fill=red!30]
\tikzstyle{func} = [rectangle, dashed, minimum width=1cm, minimum height=0.6cm,
					text centered, draw=black, fill=green!30]
\tikzstyle{class} = [rectangle, rounded corners, minimum height=0.6cm, minimum
					 width=1cm,text centered, draw=black, fill=blue!30]
\tikzstyle{block} = [rectangle, dashed, minimum width=1cm, minimum height=0.6cm,
					 text centered, draw=black, fill=white]

\tikzstyle{arrow} = [thick,->,>=stealth]
\tikzstyle{line} = [-,>=stealth]

\usepackage[compat=1.1.0]{tikz-feynman}
\usepackage{contour}

\newcommand{\vegasflow}{\texttt{VegasFlow}~}
\newcommand{\pdfflow}{\texttt{PDFFlow}~}
\newcommand{\vegasflowpdfflow}{\texttt{VegasFlow-PDFFlow}~}

\definecolor{darkgreen}{rgb}{0.0, 0.5, 0.13}
\definecolor{darkred}{rgb}{0.55, 0.0, 0.0}
\title{PDFFlow: hardware accelerating parton density access}
\ShortTitle{PDFFlow}

\author*[a,b]{Marco Rossi}
\author[a]{Stefano Carrazza}
\author[a]{Juan M. Cruz-Martinez}

\affiliation[a]{TIF Lab, Dipartimento di Fisica, Universit\`a degli Studi di 
                Milano and INFN Sezione di Milano \\
                Via Celoria 16, 20133, Milano, Italy}
\affiliation[b]{CERN openlab, Geneva 23, CH-1211, Switzerland}

\emailAdd{stefano.carrazza@mi.infn.it}
\emailAdd{juan.cruz@mi.infn.it}
\emailAdd{marco.rossi5@unimi.it}

\abstract{
	We present \texttt{PDFFlow}, a new software for fast evaluation of parton
    distribution functions (PDFs) designed for platforms with hardware
    accelerators. PDFs are essential for the calculation of particle physics
    observables through Monte Carlo simulation techniques. The evaluation of a
    generic set of PDFs for quarks and gluons at a given momentum fraction and
    energy scale requires the implementation of interpolation algorithms as
    introduced for the first time by the LHAPDF project. \pdfflow
    extends and implements these interpolation algorithms using Google's
    TensorFlow library providing the possibility to perform PDF evaluations
    taking fully advantage of multi-threading CPU and GPU setups. We benchmark
    the performance of this library on multiple scenarios relevant for the
    particle physics community.
}

\FullConference{%
  40th International Conference on High Energy physics - ICHEP2020\\
  July 28 - August 6, 2020\\
  Prague, Czech Republic (virtual meeting)
}

\note{\textit{Preprint numbers:} TIF-UNIMI-2020-31, CERN-IT-2020-001}
\begin{document}

\maketitle

\section{Introduction and motivation}
\label{sec:introduction}

Parton Distribution Functions (PDFs) are a core concept in High Energy
Physics (HEP) phenomenology: they provide a description of the parton content
of the proton and enter the computation of physical observables, like cross
sections and differential distributions. PDFs are provided by fitting
collaborations, each following different techniques and
methodologies~\cite{Ball:2017nwa, Hou:2019jgw,Harland-Lang:2014zoa}.
Nonetheless, a standard format~\cite{Whalley:2005nh} was agreed to be used by
the community: PDF measurements are presented in grids of values as functions of
momentum fraction $x$ of a parton with flavor $a$ and energy scale $Q$;
collections of such grids are usually called PDF sets.

The state of the art tool that allows access to PDF values is represented by the
LHAPDF library~\cite{Buckley:2014ana}, whose algorithms ask as input a single
query point in the $(a, x, Q)$ space and output the corresponding interpolated
value from the PDF grid. We identify the intrinsic sequential nature of the
LHAPDF tool as the entry point of our research development: physical calculations are
usually performed via Monte Carlo (MC) simulations, see
section~\ref{subsec:usage}, asking to produce a huge number of independent phase
space points; then, parallelizing such uncorrelated
computations, may lead to massive performance improvements, like
in~\cite{Campbell:2015qma}.

Moreover, in recent years the decreasing costs of hardware accelerators, such
as GPUs, TPUs and FPGAs, triggered a raising interest in the development of
software and frameworks that target these platforms to reduce the computational
burden of HEP phenomenological simulations.
\vegasflow~\cite{Carrazza:2020rdn,juan_cruz_martinez_2020_3691927} is an
excellent example of such approach: written in python, it leverages the
TensorFlow~\cite{tensorflow2015:whitepaper} framework to delegate the placement
and management of tensors on hardware, allowing the developer and the user to
focus on the actual implementation of the tool, rather than on the difficulties
that may arise from programming directly into hardware specific languages.

Following the strategy introduced by \vegasflow, we present the \pdfflow
library~\cite{juan_cruz_martinez_2020_3964191}, where
the main contribution is a novel implementation of the PDF and $\alpha_s$
interpolation algorithm used in LHAPDF. Able to run both in CPUs and GPUs,
it enables further acceleration for Monte Carlo simulations. This paper is intended
to reach users with a detailed review of \pdfflow usage and examples of real
physical use cases, along with a presentation of main
benchmark results. The paper is structured as follows, in
section~\ref{sec:pdfflow} we discuss technical implementation of the tool and
its comparison with the state of the art. Section~\ref{sec:example} overviews
\pdfflow and \vegasflow integration into real LHC physics simulations, sketching
the idea that drives this computation and presenting two complete examples.
Finally, section~\ref{sec:conclusions} wraps up our considerations and future
work directions.

\section{\pdfflow}
\label{sec:pdfflow}

\subsection{Technical Implementation}
\label{subsec:techimp}

The idea behind \pdfflow implementation is to mimic the LHAPDF interpolation
algorithms for PDFs evaluation, trying to parallelize as much as possible the
computation and exploiting TensorFlow library to be hardware agnostic, in order
to benefit from running on a wide spectrum of modern hardware. \pdfflow then, as opposite to LHAPDF, groups
together multiple points calculations, asking as inputs three arrays: a set of
flavors $\{a\}$ and two arrays of the same length, namely the fractions of
momenta $x_i$ and energy scales $Q_i$ respectively. The output of such query is an array of PDF interpolated
points $f_i$.

\begin{figure}
    \begin{minipage}{0.49\textwidth}
        \begin{tiny}
            \begin{minted}[autogobble]{python}
from pdfflow import mkPDF
pdf = mkPDF(f"{pdfset}/0")
pdf.trace() # pdf initializer

pid = [-1,21,1] # anti-down, gluon, down flavors
x = [10**i for i in range(-6,-1)] # momenta fractions
q = [10**i for i in range(1,6)] # energy scales

f = pdf.py_xfxQ(pid, x, q)



###############################################################



from pdfflow import mkPDF
pdf = mkPDF(f"{pdfset}/0")
pdf.alphas_trace() # pdf initializer

q = [10**i for i in range(1,6)] # energy scales

alphas = pdf.py_alphasQ(q)

            \end{minted}
        \end{tiny}
        \caption{Interpolation example code. \emph{Top}: PDF interpolation.
        \emph{Bottom}: $\alpha_s$ interpolation. It's good practice to call the \texttt{trace} or
        \texttt{alphas\_trace} initializers the first time a \texttt{PDF} is created.}
        \label{code:pdfflow}
    \end{minipage}
    \hfill
    \begin{minipage}{0.49\textwidth}
        \center
        \includegraphics[height=6.5cm]{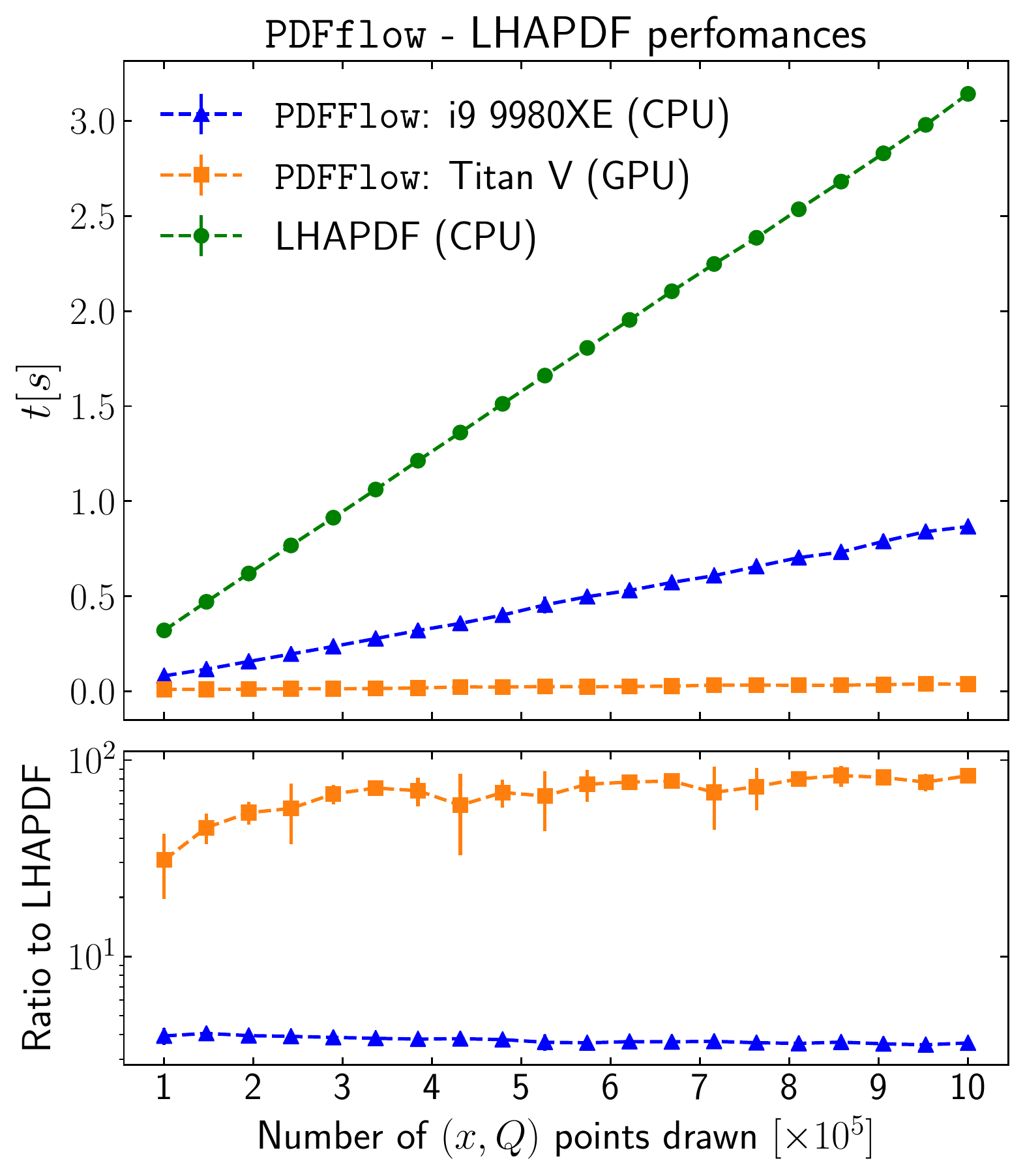}
        \caption{Performance benchmark. \emph{Top}: absolute execution time as a
        function of the query array length. \emph{Bottom}: time ratio between
        \pdfflow running on different platforms and LHAPDF.}
        \label{fig:performance}    
    \end{minipage}    
\end{figure}

\pdfflow provides a user friendly python interface, where a \texttt{PDF} object is istantiated via a \texttt{mkPDF}
function, while the actual interpolation is called by the \texttt{py\_xfxQ} \texttt{PDF} attribute
method. Figure~\ref{code:pdfflow} shows how a \pdfflow user may start an
interpolation in their code.\footnote{A complete \emph{"How to"} documentation is
available at the
\href{https://github.com/N3PDF/pdfflow/tree/master/doc/source}
{\textcolor{blue}{N3PDF/pdfflow}} repository.} Analogous to the PDF
interpolation problem is the running of the strong coupling $\alpha_s$
one. Like LHAPDF, we implement this algorithm, whose usage is reported also in
the figure.

\subsection{Benchmarks}
\label{subsec:benchmarks}

We present here benchmark results between \pdfflow \texttt{v1.0} and
{\tt LHAPDF v6.3.0} libraries. While we refer to~\cite[section~3.2]{Carrazza:2020pdf}
for the accuracy performance, claiming a perfect agreement between the two tools
outputs, here we focus on the performance comparison. In figure~\ref{fig:performance}
we show the execution time as a function of the query array length. Different
curves correspond to different hardware setups: \pdfflow provides a good speed
up while running on an Intel i9 CPU (blue curve), showing a flat ratio factor varying
between $3$x and $4$x. The behavior dramatically improves when testing our
tool on a Titan V GPU (orange curve), achieving a best improvement of two orders
of magnitude when we query $10^6$ input points. The higher number of cores that
compute parallel operations in a GPU provides tremendous improvement, even with
a non-GPU specifically optimized code.

\section{\vegasflowpdfflow usage and examples}
\label{sec:example}

\subsection{Monte Carlo simulations use cases}
\label{subsec:usage}

Monte Carlo simulations represent a natural
framework where a parallelized algorithm can leverage all its capabilities,
given that calculations involving different integration points are completely
independent from each other. In this section we show how to exploit \pdfflow and \vegasflow to compute cross
sections at hadron colliders. The problem can be cast into the following form:
\begin{wrapfloat}{figure}{r}{0.49\textwidth}
    \centering
    \begin{tikzpicture}[node distance = 1cm]
        \centering
        \small
        \node[tool] (ex)
         at (4.25,0) {Simulation Example};
             
        \node[block, text width=2.cm, minimum height=1cm, minimum width=2.cm]
         (vflow) at ($(ex)+(-2.1,-1.4)$) {\texttt{VegasFlow:\\random gen}};
             
        \node[block, text width=1.5cm, minimum height=1cm, minimum width=1.5cm]
         (pflow) at ($(ex)+(0,-1.4)$) {\texttt{PDFFlow:\\PDF eval}};

        \node[func, text width=2cm, minimum width=2cm] (kin)
         at ($(vflow)+ (0,-1.5)$) {\texttt{Event\\kinematics}};

        \node[func, text width=1.5cm, minimum width=1.5cm] (ME)
         at ($(kin)+ (0,-1.5)$) {\texttt{Matrix\\element}};
             
        \node[func, text width=2.cm, minimum width=2.cm] (lumi)
         at ($(pflow)+ (0,-3)$) {\texttt{Luminosity}};

        \node[func, text width=1.8cm, minimum width=1.8cm] (PS)
         at ($(lumi)+(2.2,0)$) {\texttt{Flux,\\PS factors}};

        \node[block, text width=2.1cm, minimum height=1cm, minimum width=2.1cm]
         (integration) at ($(lumi)-(0,1.6)$) {\texttt{VegasFlow:\\integration}};

        \draw [decorate,decoration={brace,amplitude=10pt}]
         ($(ex)-(3.5,0.8)$) -- ++(right:7);

        \draw[arrow] (vflow) -- (kin);
        \draw[arrow] (pflow) -- (lumi);

        \draw[arrow] (kin) -- (ME);

        \draw[line, thick] ($(ME)+(0,0.75)$) -- ($(PS)+(0,0.75)$);
        \draw[arrow] ($(PS)+(0,0.75)$) -- (PS);

        \draw[line,thick] (ME) -- ++(down:0.75);
        \draw[line,thick] (PS) -- ++(down:0.75);
        \draw[line, thick] ($(ME)-(0,0.75)$) -- ($(PS)-(0,0.75)$);

        \draw[arrow] (lumi) -- ++(down:1.1);

        \draw[dashed, fill=green!30, fill opacity=0.2]
            ($(PS.south east)+(0.05,-0.15)$) rectangle
            ($(kin.north west)+(-0.05,0.2)$);
        \node at ($(PS)+(0,1.8)$) {\texttt{Integrand}};
    \end{tikzpicture}
    \caption{\pdfflow flowchart. Blocks are color-coded as following: white for
    tools implemented algorithms, green for required example-dependent
    user-defined functions.}
    \label{fig:flowchart}
\end{wrapfloat}
\begin{equation}
    \sigma_{X} = \int dx_1 dx_2 \frac{\Phi}{\mathrm{F}} \sum_{a,b}
    \Big( \mathcal{L}_{a,b} \lvert \mathcal{M}_{a,b \to X} \rvert^2 \Big)
    \label{sigma:rewritten}
\end{equation}
where the luminosity $\mathcal{L}$ accounts for the PDFs, the matrix element
$\mathcal{M}$ describes the hard interaction, while $\mathrm{F}$ and $\Phi$ are
the flux and phase space factors, respectively. This integral can be
evaluated numerically exploiting \vegasflow and \pdfflow tools. The general
implementation scheme is presented in figure~\ref{fig:flowchart}.

The user has to define an \texttt{Integrand} function that will be called by
\vegasflow at integration time, returning the
process cross section and its uncertainty. This function encodes the
process-specific physical factors we highlighted in
equation~\eqref{sigma:rewritten}: it must take as input an array of random
numbers, provided by \vegasflow, and output the function to be integrated,
evaluated at those points. In order to build this function, the user must 
convert the plain random numbers into physical variables, like $x_{1,2}$, that
describe the event kinematics. Note that, in this step, the jacobian factor of
the trasformation must be taken into account to obtain the correct result. After
having reconstructed the kinematics, one must compute the matrix element plus
the flux and phase space factors. These ingredients in turn must be combined
with the proper luminosity function, computed evaluating the PDFs at the correct
$(x,Q^2)$ points by means of \pdfflow \texttt{xfxQ2}.

\subsection{Single top @ LO}
\label{subsec:singletop}

\begin{figure}[t]
    \centering
    \hfill
    
\end{figure}
As a basic example, we consider the single-top production at leading order (LO)
in proton-proton collisions through the exchange of a $W$ boson in the
$t-$channel. 
\begin{figure}
    \center
    \begin{minipage}[t]{0.49\textwidth}
    \begin{tiny}
        \begin{minted}[autogobble]{python}
import tensorflow as tf
from vegasflow import vegas_wrapper
from pdfflow import mkPDF
...
p = mkPDF(pdfset, DIRNAME) # build PDF
...
@tf.function
def get_x1x2(xarr):
    """ Maps vegas random points to x1,x2
        Returns: partonic center of mass energy,
                 jacobian, x1, x2 """
    ...
    return shat, jac, x1, x2

@tf.function
def make_event(xarr):
    """ Returns the kinematics:
        psw, p0, p1, p2, p3, p4, x1, x2 """
    shat, jac, x1, x2 = get_x1x2(xarr) # transformation
    return build_kinematics(shat, jac, x1, x2)    
        \end{minted}
    \end{tiny}
    \end{minipage}
    \hfill
    \begin{minipage}[t]{0.49\textwidth}
    \begin{tiny}
        \begin{minted}[autogobble]{python}
@tf.function
def build_luminosity(x1,x2):
  q2s = tf.ones_like(x1)*mt2 # eval PDFs at top mass squared
  p5x1 = pdf.xfxQ2([5], x1, q2s) # bottom PDFs
  pNx2 = pdf.xfxQ2([2, 4, -1, -3], x2, q2s) # u, c, dx, bx PDFs
  lumis = muliplyPDFs_and sum(p5x1, pNx2)
  return lumis / x1 / x2

@tf.function
def singletop(xarr):
  """ Input: xarr, vegas random points"""
  psw, flux, p0, p1, p2, p3, x1, x2 = make_event(xarr)
  wgts = evaluate_matrix_element_square(p0, p1, p2, p3)
  lumi = build_luminosity(x1,x2)
  lumi_ME = combine_lumi_ME(lumi, wgts)
  return psw * flux * lumi_ME
    
r = vegas_wrapper(singletop, dim, n_iter, ncalls, compilable=True)
        \end{minted}
    \end{tiny}
    \end{minipage}
    \caption{Code for single-top production in $t-$channel at LO:
    \texttt{singletop} function evaluates the integrand and is passed
    as a parameter to \texttt{vegas\_wrapper} function; \pdfflow enters in the
    luminosity computation.}
    \label{code:singletop}
\end{figure}
Figure~\ref{code:singletop} instructs the fundamental steps
required to build the example code\footnote{The entire code is
publicly available inside the
\href{https://github.com/N3PDF/pdfflow/blob/master/benchmarks/singletop_lo.py}
{\textcolor{blue}{N3PDF/pdfflow}} repository.}.

In figure~\ref{fig:singletop} we show the performance comparison between
\vegasflowpdfflow, running on multiple hardware setups, against fixed order
calculation at LO with MG5\_aMC@NLO~\cite{Alwall:2014hca} software.
We stop both the integrations when the total number of generated events allows
to achieve a precision better than $2 \cdot 10^{-3}$\SI{}{\pico\barn} (relative
error of $4 \cdot 10^{-5}$) on the output cross section. The bar chart
underlines how our parallelized approach grants a great speed up against the
baseline algorithm. The picture is even more extreme when running on GPU.

\subsection{VFH @ NLO}
\label{subsec:vbf}

Monte Carlo calculations become exponentially more and more
expensive in terms of computational and human efforts at higher orders in
perturbation theory. This added complexity provides a perfect benchmarking
ground for assessing the GPU acceleration of our models. In particular, we test
\vegasflowpdfflow setup against a simplified version of the existing
Fortran $95$ NNLOJET~\cite{Cruz-Martinez:2018rod} implementation of the vector
boson fusion for Higgs production (VFH) at NLO. In this picture we consider only
the quark initiated W-boson mediated process, eventually with a gluon emitted
from any of the quarks at NLO.

\begin{figure}
    \centering
    \begin{minipage}[t]{0.49\textwidth}
        \centering
        \includegraphics[width=0.95\textwidth]{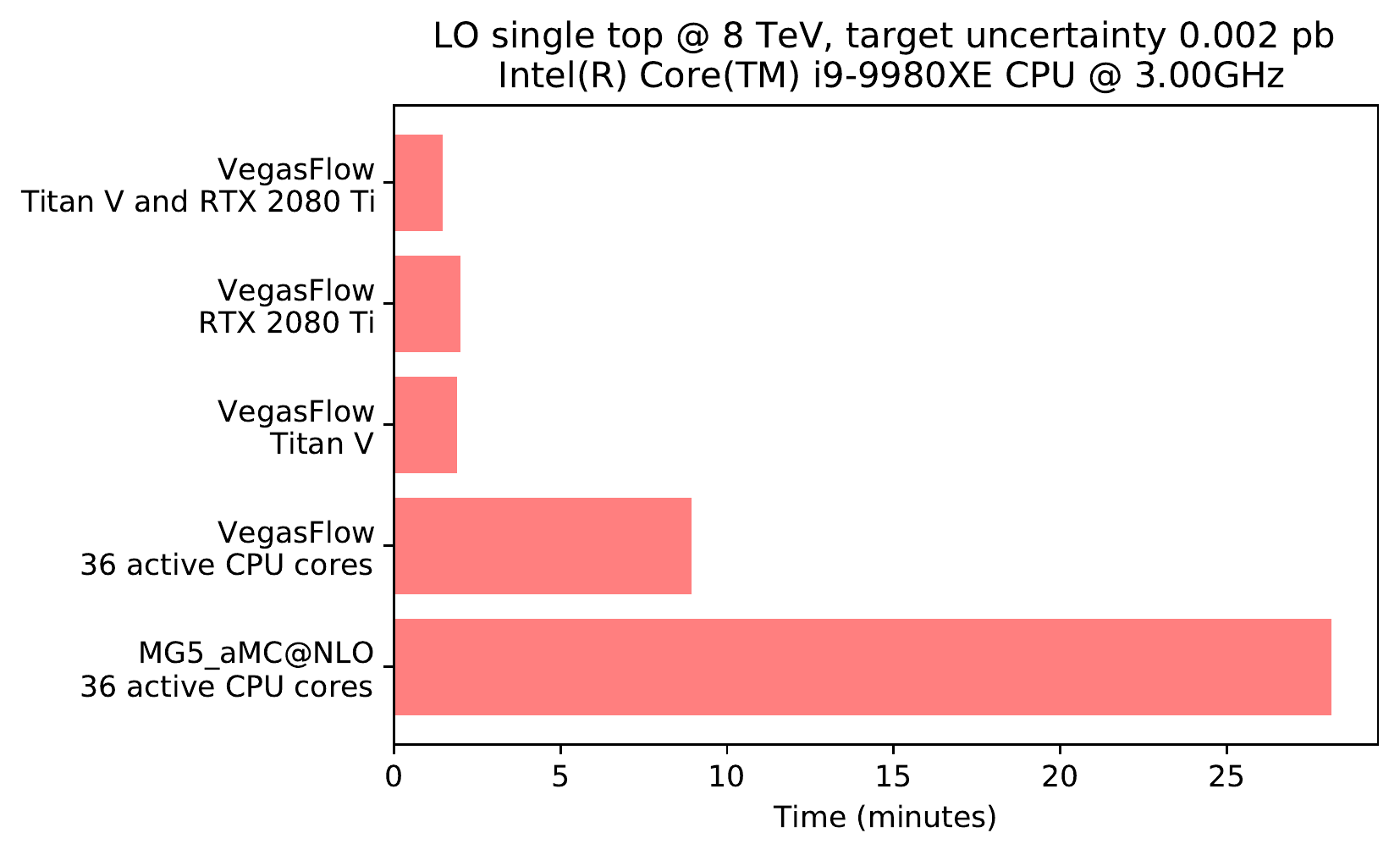}
        \caption{Total integration time comparison between \vegasflowpdfflow 
        and MG5\_aMC@NLO at LO.}
        \label{fig:singletop}
    \end{minipage}
    \hfill
    \begin{minipage}[t]{0.49\textwidth}
        \centering
        \includegraphics[height=5cm]{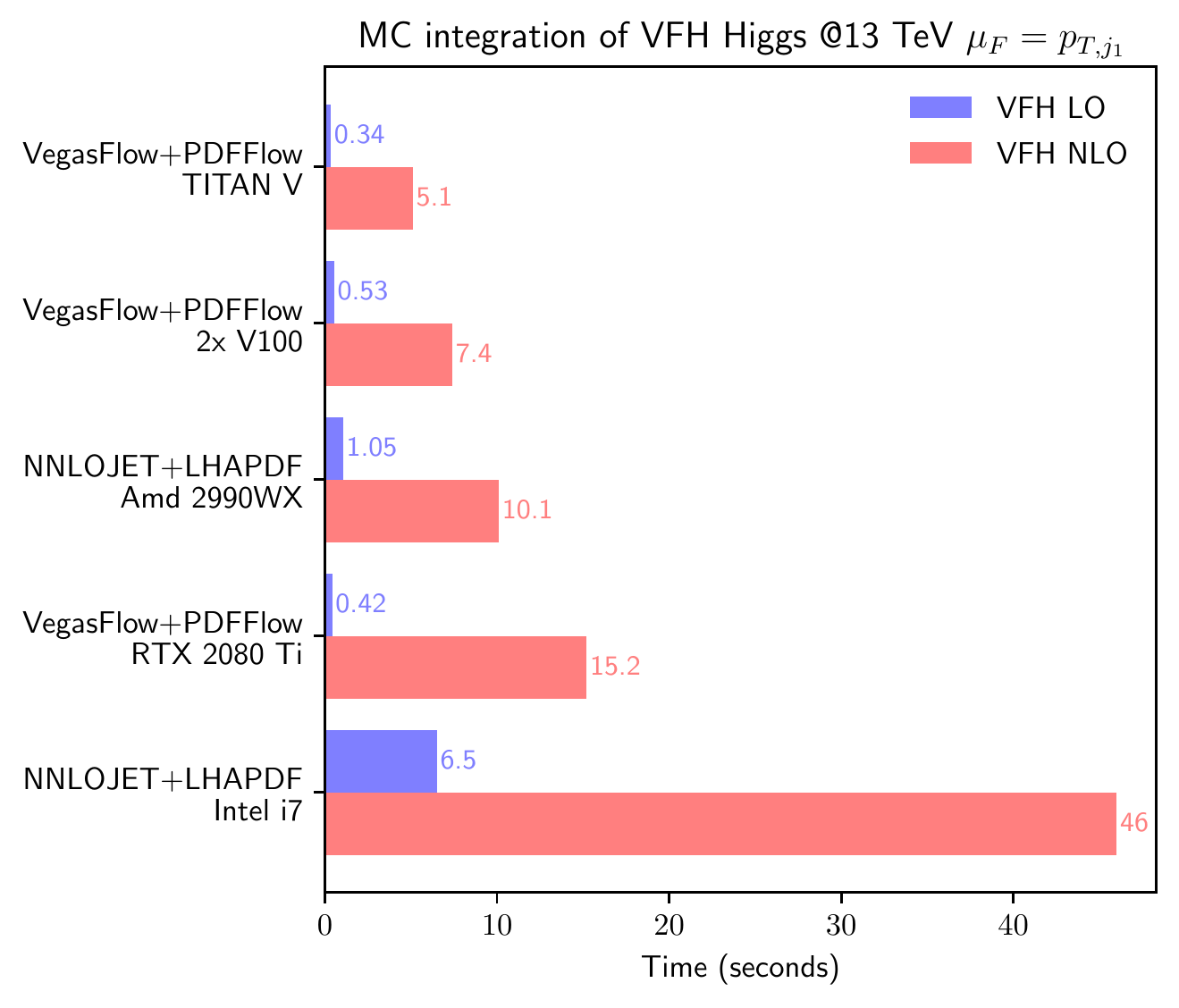}
        \caption{\small Time per iteration comparison of \vegasflowpdfflow and
        NNLOJET at NLO.}
        \label{fig:vfh}
    \end{minipage}
\end{figure}

We collect performance results in figure~\ref{fig:vfh}, evaluating algorithms on
multiple platforms. The stopping
criteria for the integrators is set to per-mille precision at LO and percent at
NLO in 10 iterations. Although NNLOJET code is highly optimized for CPU and
CPU-cluster usage, we observe that our tool, with little to no specific GPU
optimization, outperforms the baseline both for LO and NLO accuracy.

\section{Conclusions}
\label{sec:conclusions}

Porting PDFs to GPU is an essential step in order to accelerate Monte Carlo
simulation by granting to the HEP community the ability to implement with
simplicity particle physics processes without having to know about the
technicalities or the difficulties of their implementation on multi-threading
systems or the data placement and memory management that GPU and multi-GPUs
computing requires.

\pdfflow is designed to work in synergy with the LHAPDF library, therefore
it uses exactly the same PDF data folder structure, and interpolation algorithms
for the PDF and $\alpha_s$ determination. While the current release of \pdfflow
has only been tested in GPUs and CPUs showing great performance improvements,
we believe that investigation about new hardware accelerators such as Field
Programmable Gate Arrays (FPGA) and Tensor Processing Units (TPUs) could provide
even more impressive results in terms of performance and power consumption.

\acknowledgments
The authors are supported by the European Research Council under the European
Union's Horizon 2020 research and innovation Programme (grant agreement number
740006) and by the UNIMI Linea 2A grant ``New hardware for HEP''.

\scriptsize
\bibliographystyle{../JHEP}
\bibliography{../blbl}
\end{document}